# Bell's measure and implementing quantum Fourier transform with orbital angular momentum of classical light

Xinbing Song,[1*] Yifan Sun,[1,2*] Hongwei Qin,[1], Pengyun Li[1] and Xiangdong Zhang[1+]

[1]School of Physics, Beijing Institute of Technology, 100081, Beijing, China
[2]Department of Physics, Beijing Normal University, Beijing 100875, China

## Abstract

We perform Bell's measurement for the non-separable correlation between polarization and orbital angular momentum from the same classical vortex beam. The violation of Bell's inequality for such a non-separable classical correlation has been demonstrated experimentally. Based on the classical vortex beam and nonquantum entanglement between the polarization and orbital angular momentum, the Hadamard gates and conditional phase gates have been designed. Furthermore, a quantum Fourier transform has been implemented experimentally, which is the crucial final step in Shor's algorithm.

*These authors contributed equally to this work.
[+]To whom correspondence should be addressed. E-mail: zhangxd@bit.edu.cn

# I. INTRODUCTION

Bell's measure is commonly used in tests of quantum non-locality, it has attracted much attention in the last years due to the possibility of ruling out classical hidden-variable theories [1-5]. Recently, it has been demonstrated that Bell's measure can also be used as a quantitative tool in classical optical coherence [6]. Non-separable correlations among two or more different degrees of freedom from the same classical optical beam, have been discussed [6-19]. The violation of Bell's inequality for such a non-separable correlation has been demonstrated experimentally [6-9]. Such a non-separable classical correlation is called "nonquantum entanglement" or "classical entanglement" [10-19]. It has been applied to resolve basic issues in polarization optics [10], simulate quantum walks, et.al. [12].

So far, the classical entanglements between polarization and some spatial modes such as spatial parity and Hermite modes, have been demonstrated experimentally in polarized beams of light [6-8]. On the other hand, vortex beams with various orbital angular momentum (OAM) have been experimentally realized in the optical domain [20-23]. The possibility of encoding large amounts of information in vortex beams due to the absence of an upper limit has raised the prospects of their applicability in quantum information processing tasks, such as computation and cryptography [22, 23]. Although Bell-like inequality for the spin-orbit separability of a laser beam has been discussed [7], direct Bell's measurement for the non-separable correlation between polarization and OAM from the same classical vortex beam has not been done.

In this work, we perform direct Bell's measurement between polarization and OAM from the same classical vortex beam, and explore classical entangled properties between them. Based on such a nonquantum entanglement, we implement the quantum Fourier transform (QFT), which is the crucial final step in Shor's algorithm [24-30]. Comparing with the quantum realization, the classical implementation of QFT exhibits many advantages. It is not only easier to implement, the measurement efficiency is also high. Thus, we look forward to our study can provide an important reference for the classical and quantum information processes.

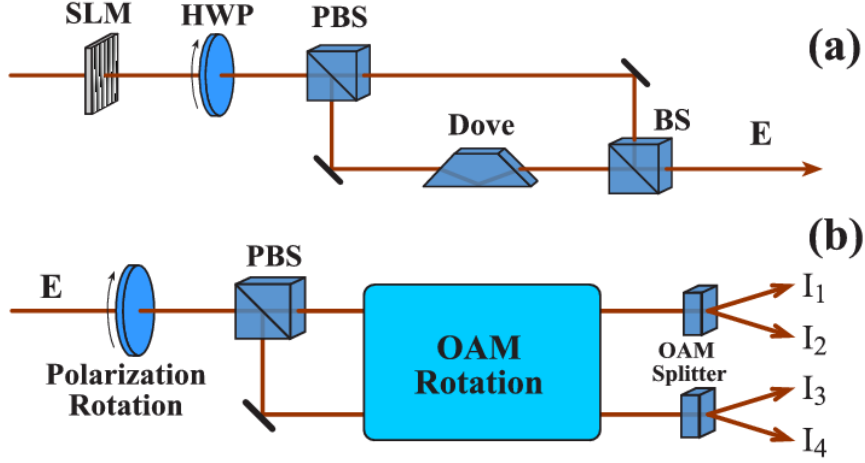

**Fig.1. (a) The preparation of classical entangled states for polarization and orbital angular momentum. (b) Experimental setup for CHSH-type Bell's measurement for the correlation between polarization and orbital angular momentum. SLM is spatial light modulator, HWP is half-wave plate，PBS is polarizing beam splitter, BS is 50/50 beam splitter, Dove is Dove prism, and OAM represent orbital angular momentum.**

## II. BELL'S MEASUREMENT FOR THE CORRELATION BETWEEN POLARIZATION AND ORBITAL ANGULAR MOMENTUM

We consider a light beam being manipulated with a spatial light modulator (SLM) pass through a half-wave plate (HWP) and a polarizing beam splitter (PBS) as shown in Fig.1(a). After a PBS, the light beam is divided into two intensity equaled parts for two paths with horizontal ($\hat{h}$) and vertical ($\hat{v}$) polarization, respectively. In the path with $\hat{v}$ polarization, a Dove prism is introduced. Then, two polarized vortex beams in two paths combine by a beam splitter (BS), the output of polarized vortex beam can be expressed as

$$\mathbf{E} = LG_{pl}\hat{h} + LG_{p-l}\hat{v}, \qquad (1)$$

where $LG_{pl}$ represents the $lth$ Laguerre functions and p is the radial node number. If the horizontal and vertical polarization components of the vortex beam are described by a slightly modified version of the familiar bra-ket notation of quantum mechanics, $|H)$ and $|V)$, the OAM of $LG_{pl}$ is expressed by $|\pm l)$, a polarized vortex beam can be described by the ket notation

$$|E) = \frac{1}{\sqrt{2}}\left[|\text{H},+l) + |\text{V},-l)\right]. \qquad (2)$$

This representation for the polarized vortex beam is formally equivalent (isomorphic) to a Bell state of two polarized qubits. Hence, the polarization and OAM may be treated as two qubits that are classically entangled. Such an entanglement is realized from a single light beam (local entanglement). The problem is whether or not such an entanglement relation can be demonstrated by Bell's measurements.

In order to answer such a problem, an experiment was designed as shown in Fig.1(b). A polarized vortex beam passes through an HWP and a PBS to split into two beams, then, they are manipulated with OAM rotation system and splitter devices to become four beams. The output intensities of four vortex beams are marked in Fig.1(b) by $I_1$, $I_2$, $I_3$ and $I_4$, respectively. In order to perform the Clauser–Horne–Shimony–Holt (CHSH) Bell's measurement, we define the following correlation function [6]:

$$C(\theta,\phi) = P_{H,+l}(\theta,\phi) - P_{H,-l}(\theta,\phi) - P_{V,+l}(\theta,\phi) + P_{V,-l}(\theta,\phi), \qquad (3)$$

where $\theta$ and $\phi$ represent polarization and OAM rotated angles in the paths, respectively. The $P_{P,O}(\theta,\phi)$ ($P = H(V)$ for the polarization and $O = \pm l$ for the OAM number) are normalized probabilities of the states on the certain measurement basis, which can be obtained through the intensity measurements in the experiments, that is $P_{H,+l}(\theta,\phi) = I_1/I$, $P_{H,-l}(\theta,\phi) = I_2/I$, $P_{V,+l}(\theta,\phi) = I_3/I$, $P_{V,-l}(\theta,\phi) = I_4/I$ and $I = I_1 + I_2 + I_3 + I_4$. After we have obtained $C(\theta,\phi)$, the CHSH measurement is

$$B = |C(\theta,\phi) + C(\theta'',\phi) + C(\theta,\phi'') - C(\theta'',\phi'')|. \qquad (4)$$

Figure 2 (a) and (b) shows experimental results for the correlation functions $C(\theta,\phi)$ from the same vortex beam with $l = \pm 1$ as a function of $\theta$ and $\phi$. Here OAM rotation has been realized by Dove prism (and two conventional $\pi/2$ astigmatic mode converters [31]) and the OAM splitter has been finished by the SLM as shown in the experimental setup on top of Fig.2. A continuous laser with wavelength $632.8\,nm$ is used.

The circle dots and solid lines represent the experimental measurements and theoretical results, respectively. Here, the theoretical results are normalized by the experimental data. It can be seen that the experimental results are in good agreement with the theoretical calculations in change characteristics. If we take $\theta = -\pi/8$, $\theta'' = \pi/8$, $\phi = 0$ and $\phi'' = -\pi/4$, $B = 2.407 \pm 0.035$ can be obtained. The theoretical result for $B$ is $2\sqrt{2}$. There are two main reasons for such a difference. On the one hand, it comes from the OAM rotating module, that is, it is not perfect for the operation effect of cylindrical lens and Dove prism on the OAM rotation. On the other hand, it is originated from imprecise measurements of output intensities of four vortex beams. Although there are some loss, the presented experiment yields the strongest violation of Bell's inequalities.

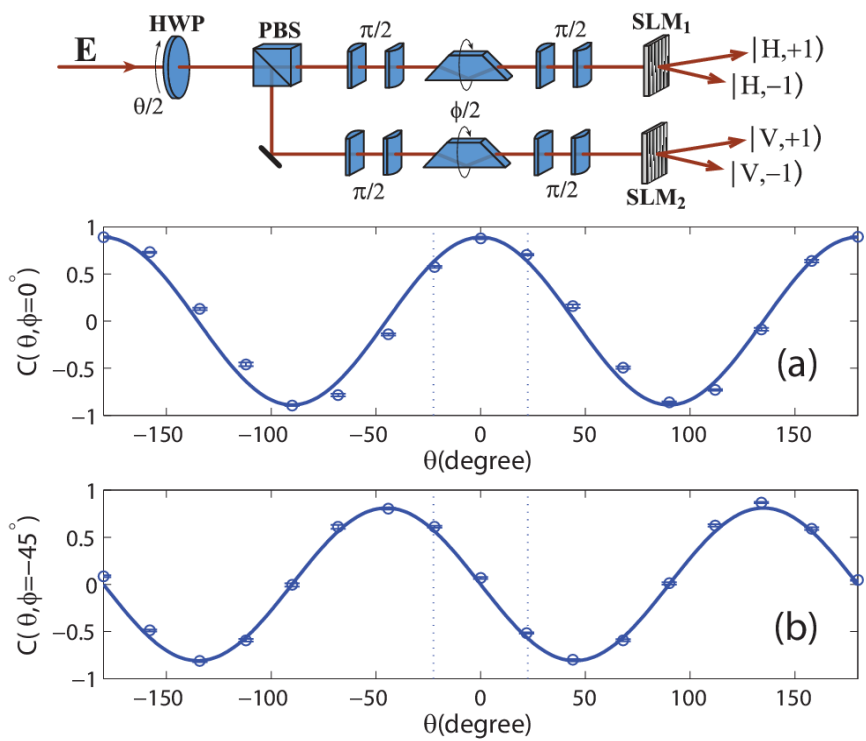

**Fig.2 The correlation functions $C(\theta,\phi)$ as a function of polarization rotated angle $\theta$ at $\phi = 0^0$ (a) and $\phi = 45^0$ (b). The circle dots and solid lines represent the experimental and theoretical results, respectively. The dashed lines mark the values of $\theta$ to achieve the maximum violations of Bell inequalities. Experimental setup for the CHSH-type Bell's measurement for the correlation between polarization and orbital angular momentum are shown on top of the figure.**

The experimental results in Fig.2 only exhibit the classical correlation between the polarization and OAM with $l = \pm 1$. In principle, the classical correlation between the polarization and OAM with any mode number can also be tested through such an experimental design. However, it is very difficult to operate in this scheme. Thus, we take another scheme, which the detailed design has been given in the Appendix. The experimental results for $l = \pm 2$ are plotted as circle dots in Fig.3. The solid lines are theoretical results, which are normalized by the experimental data. The agreements between the experimental measurements and theoretical results in change characteristics are observed again. If we take $\phi' = \pm 22.5°$ $\theta' = 0°$ and $-45°$, B=2.101±0.028 can be obtained. Here the theoretical result for B is still $2\sqrt{2}$. Comparing the results in Fig.3 with those for $l = \pm 1$ in Fig.2, we find that the experimental results still yield the violation of Bell's inequalities although the loss in the experimental process increases for the present case with $l = \pm 2$. With the increase of $l$, it becomes more difficult to obtain the measured results with B>2 because it requires more accurate measurement of intensity and precise operation on optical elements. However, the fact for violating the Bell's inequalities can be confirmed.

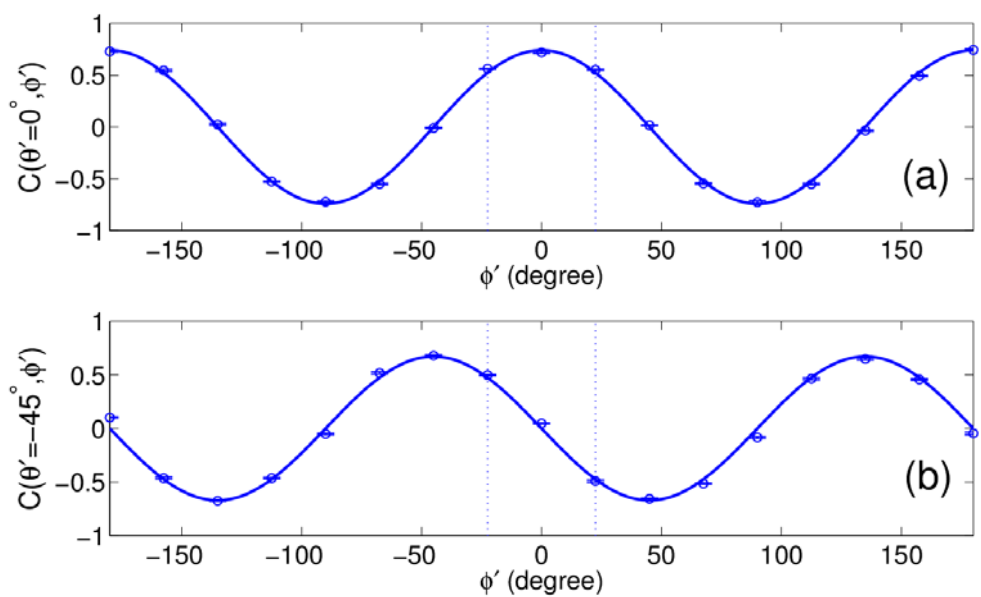

**Fig. 3. The correlation functions $C(\theta', \phi')$ as a function of polarization rotated angle $\theta'$ at $0°$ (a) and $-45°$ (b) for $l = 2$. The circle dots and solid lines represent the experimental and theoretical results, respectively. Here $\phi' = \pm 22.5°$, $\theta'$ and $\phi'$ still represent polarization**

and OAM rotated angles, but their values are different from $\theta$ and $\phi$ in Fig.2 (see Appendix).

This means that the local classical entanglement state between the polarization and OAM has been demonstrated. The problem is whether or not such a classical entanglement state can be applied to perform quantum information process such as quantum algorithm. Because the QFT is the crucial final step in some quantum algorithms such as Shor's algorithm, in the following we explore possibility to realize QFT by using the local classical entanglement state between the polarization and OAM.

## III. QUANTUM FOURIER TRANSFORM BASED ON THE CLASSICAL VORTEX BEAM

The QFT is a basis transformation in an N-state space that transforms the state $|k\rangle$ according to

$$|k\rangle \to \frac{1}{\sqrt{N}} \sum_{j=0}^{N-1} e^{i2\pi jk/N} |j\rangle, \qquad (5)$$

where $|k\rangle, |j\rangle \in \{|n\rangle\}_N$ and $\{|n\rangle\}_N$ is a set of complete orthogonal basis vectors with N dimension, k and j represent an integer ranging from 0 to N-1. So far, the QFT has been demonstrated experimentally by using quantum Hadamard gate and conditional phase gates [32]. The main advantage of QFT against the classical Fourier transformation is higher calculation efficiency, which is originated from the quantum correlation.

The above investigations have shown that classical entanglement states can exhibit similar correlation properties with the quantum correlation. If we take classical states $|k)$ and $|j)$ instead of quantum states $|k\rangle$ and $|j\rangle$, the similar transformation $|k) \to \frac{1}{\sqrt{N}} \sum_{j=0}^{N-1} e^{i2\pi jk/N} |j)$ can be completed by using the local classical entanglement states between the polarization and OAM. In the following, we take two-qubit as an example to demonstrate such a process. The polarization freedom is marked as the first qubit, that is $|0) \to |H)$ and $|1) \to |V)$, the OAM is marked as the second qubit, $|0) \to |+1)$ and $|1) \to |-1)$, then four Bell's states are expressed as:

$$|\psi_\pm^C\rangle = \frac{1}{\sqrt{2}}[|H,-1\rangle \pm |V,+1\rangle],$$
$$|\varphi_\pm^C\rangle = \frac{1}{\sqrt{2}}[|H,+1\rangle \pm |V,-1\rangle]. \qquad (6)$$

In order to realize QFT of four Bell states, we present an experimental setup as shown in Fig.4. It consists of two Hadamard gates and one conditional phase gate. The first Hadamard gate is for the OAM, which has been realized by using Dove prism with fixed rotated angle $67.5^0$. The second Hadamard gate is for the polarization, which can be realized by a HWP. The conditional phase gate consists of one spiral phase plate (SPP) (RPC photonics, VPP-m633) and one SLM. The function of SPP is to change the order of QAM, that is $|-1\rangle \to |0\rangle$ and $|1\rangle \to |2\rangle$. Let four Bell states pass through such an experimental setup, the output states are:

$$QFT\{|\psi_+^C\rangle\} = \frac{1}{2\sqrt{2}}\left[2|H,+1\rangle - (1-i)|H,-1\rangle - (1+i)|V,-1\rangle\right]$$
$$QFT\{|\psi_-^C\rangle\} = \frac{1}{2\sqrt{2}}\left[-2|V,+1\rangle + (1+i)|H,-1\rangle + (1-i)|V,-1\rangle\right]$$
$$QFT\{|\varphi_+^C\rangle\} = \frac{1}{2\sqrt{2}}\left[2|H,+1\rangle + (1-i)|H,-1\rangle + (1+i)|V,-1\rangle\right] \qquad (7)$$
$$QFT\{|\varphi_-^C\rangle\} = \frac{1}{2\sqrt{2}}\left[2|V,+1\rangle + (1+i)|H,-1\rangle + (1-i)|V,-1\rangle\right]$$

In order to obtain the information of output states, we measure the intensity by using PBS and diaphragm as shown in Fig.4. That is to say, measure the output intensities for four basis vectors $|H,+1\rangle$, $|H,-1\rangle$, $|V,+1\rangle$ and $|V,-1\rangle$. From the conditional phase gate, four basis vectors become $|H,+2\rangle$, $|H,0\rangle$, $|V,+2\rangle$ and $|V,0\rangle$, respectively. After they pass through a PBS, the field intensities with different OAM can be obtained by shading light in space. They should be corresponded to the modular squares of coefficients for four basis vectors, if the QFT has been realized.

In Fig.5, we present the comparison between the theoretical results and experimental measurements. Figure 5 (a), (b), (c) and (d) show output ratios of different basis vectors for four kinds of input states, respectively. Here the output ratio represents the ratio of measured intensity for each basis vector and total output intensity. The blue bars represent theoretical results, and red bars are experimental results. Comparing them, we find that the agreements between the theoretical results and experimental measurements are very well. This means that the QFT has been realized by our experimental setup. We would like to point out that the above results are only for the classical vortex beam with $l = \pm 1$, the QFT can be realized by using the classical vortex beam with any mode number if the more ideal optical elements and method can be provided.

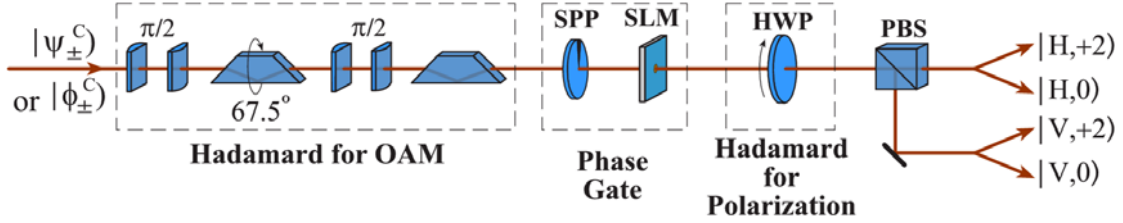

Fig. 4. Experimental setup for the quantum Fourier transform (QFT) by using the local classical entanglement state between the polarization and orbital angular momentum. The QFT is composed of two Hadamard transforms and one conditional phase gate. The SPP represents the spiral phase plate.

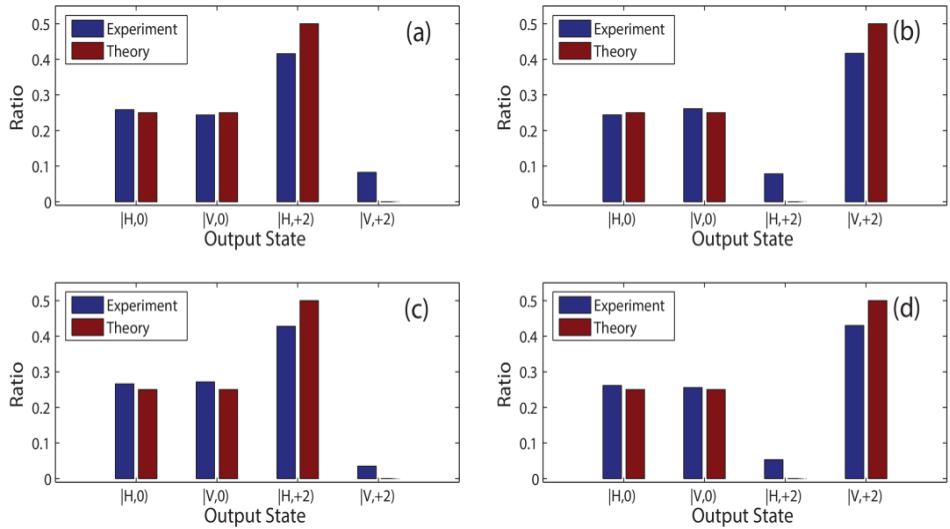

Fig. 5 Comparison between experimental results and theoretical calculations for the QFT. (a)-(d) exhibit output ratios between measured intensities for four output states ($|H,+2\rangle$, $|H,0\rangle$, $|V,+2\rangle$ and $|V,0\rangle$ corresponding to $|H,+1\rangle$, $|H,-1\rangle$, $|V,+1\rangle$, and $|V,-1\rangle$) and total intensity for four kinds of input states ($|\psi_\pm^C\rangle$ and $|\varphi_\pm^C\rangle$), respectively. The blue bars represent theoretical results, and red bars are experimental results. (a) corresponds to $|\varphi_+^C\rangle$; (b) to $|\varphi_-^C\rangle$; (c) to $|\psi_+^C\rangle$; (d) to $|\psi_-^C\rangle$.

## IV. CONCLUSION

In summary, Bell's measurement for the non-separable correlation between the polarization and OAM from the same classical vortex beam has been performed experimentally. The violations of Bell's inequalities for the non-separable classical correlations with various OAM have been demonstrated experimentally. Based on the nonquantum entanglement between the polarization and OAM in the classical vortex beam, the Hadamard gates and conditional phase gates have been designed, a QFT has been implemented experimentally. Such an implementation of QFT exhibits many advantages in comparing with usual quantum realizations. For example, it is not only easier to implement, the measurement efficiency is also high. Moreover, it sheds light on the new concept of nonquantum entanglement. The expansion of the present work to simulate quantum computing or perform other quantum information processes is a task for future research.

## ACKNOWLEDGMENTS

This work was supported by the National Natural Science Foundation of China (Grant No. 11274042 and 61421001).

## APPENDIX

In order to perform Bell's measurement for the correlation between the polarization and OAM with $|l|>1$ more efficient, we take the experimental setup as shown in Fig.A1, which is similar to that in Ref.[33]. In contrast to the scheme in Fig.2, in the present scheme the measurement basis for the OAM has been generated by controlling the polarization and the projection measurement can be realized by using the interference method. Then, the output intensities $I_1^{'}$, $I_2^{'}$ and $I_T$ as marked in Fig.A1 can be obtained. The calculated processes for $I_1^{'}$, $I_2^{'}$ and $I_T$ are given in the following:

The Jones matrix for the optical element group consisting of two HWPs and one PBS can be expressed as $\begin{pmatrix} \cos^2\theta' & \cos\theta'\sin\theta' \\ \cos\theta'\sin\theta' & \sin^2\theta' \end{pmatrix}$, where the fast axis direction for the HWP is taken as $\theta'/2$ and the transmission light from the PBS is in the horizontal polarization. For convenience, the input state can be expressed as: $|E\rangle = \frac{1}{\sqrt{2}}\begin{pmatrix}|+l\rangle \\ |-l\rangle\end{pmatrix}$, where the first line and the second line of the column matrix correspond to $|H\rangle$ and $|V\rangle$, respectively. From Fig.A1, the $I_1'$ can be calculated in the following process:

$$|E\rangle \xrightarrow{BS_0 \text{ Transmition}} \frac{1}{2}\begin{pmatrix}|+l\rangle \\ |-l\rangle\end{pmatrix} \xrightarrow{\text{HWP}(\theta'/2) \text{ +PBS+HWP}(\theta'/2)} \frac{1}{2}\begin{pmatrix}\cos^2\theta'|+l\rangle + \cos\theta'\sin\theta'|-l\rangle \\ \cos\theta'\sin\theta'|+l\rangle + \sin^2\theta'|-l\rangle\end{pmatrix}$$

$$\xrightarrow{BS_1 \text{ Transmition}} \frac{1}{2\sqrt{2}}\begin{pmatrix}\cos^2\theta'|+l\rangle + \cos\theta'\sin\theta'|-l\rangle \\ \cos\theta'\sin\theta'|+l\rangle + \sin^2\theta'|-l\rangle\end{pmatrix}. \quad (A1)$$

Considering the orthogonality of OAM modes, $\langle m|n\rangle = \delta_{mn}$, we can obtain

$$I_1' = \left|\frac{1}{2\sqrt{2}}\begin{pmatrix}\cos^2\theta'|+l\rangle + \cos\theta'\sin\theta'|-l\rangle \\ \cos\theta'\sin\theta'|+l\rangle + \sin^2\theta'|-l\rangle\end{pmatrix}\right|^2 = \frac{1}{8}. \quad (A2)$$

For the $I_2'$, we have

$$|E\rangle \xrightarrow{BS_0 \text{ Reflection}} \frac{1}{2}\begin{pmatrix}|+l\rangle \\ |-l\rangle\end{pmatrix} \xrightarrow{\text{HWP}(\phi'/2) \text{ +PBS+HWP}(\phi'/2)} \frac{1}{2}\begin{pmatrix}\cos^2\phi'|+l\rangle + \cos\phi'\sin\phi'|-l\rangle \\ \cos\phi'\sin\phi'|+l\rangle + \sin^2\phi'|-l\rangle\end{pmatrix}$$

$$\xrightarrow{\text{HWP}(\theta'/2) \text{ +PBS+HWP}(\theta'/2)}$$

$$\frac{1}{2}\begin{pmatrix}\cos^2\theta'\left[\cos^2\phi'|+l\rangle + \cos\phi'\sin\phi'|-l\rangle\right] + \cos\theta'\sin\theta'\left[\cos\phi'\sin\phi'|+l\rangle + \sin^2\phi'|-l\rangle\right] \\ \cos\theta'\sin\theta'\left[\cos^2\phi'|+l\rangle + \cos\phi'\sin\phi'|-l\rangle\right] + \sin^2\theta'\left[\cos\phi'\sin\phi'|+l\rangle + \sin^2\phi'|-l\rangle\right]\end{pmatrix}$$

$$\xrightarrow{BS_2 \text{ Reflection}} \frac{1}{2\sqrt{2}}\begin{pmatrix}-\cos(\theta'-\phi')\cos\theta'\left[\cos\phi'|-l\rangle + \sin\phi'|+l\rangle\right] \\ \cos(\theta'-\phi')\sin\theta'\left[\cos\phi'|-l\rangle + \sin\phi'|+l\rangle\right]\end{pmatrix}. \quad (A3)$$

Here a $\pi$ phase has been added in the horizontal polarized mode and the reversal has been happened for the OAM modes by the reflection, that is $|H\rangle \to -|H\rangle$ and $|+l\rangle \rightleftarrows |-l\rangle$. Then

$$I_2' = \left| \frac{1}{2\sqrt{2}} \begin{pmatrix} -\cos(\theta'-\phi')\cos\theta'[\cos\phi'|-l\rangle + \sin\phi'|+l\rangle] \\ \cos(\theta'-\phi')\sin\theta'[\cos\phi'|-l\rangle + \sin\phi'|+l\rangle] \end{pmatrix} \right|^2 = \frac{1}{8}\cos^2(\theta'-\phi'). \quad (A4)$$

The $I_T$ represents the output field intensity from the $BS_3$, and the input field comes from the reflection from the $BS_1$ and the transmission from the $BS_2$. The reflection field from the $BS_1$ is expressed as: $\frac{1}{2\sqrt{2}} \begin{pmatrix} -\cos^2\theta'|-l\rangle - \cos\theta'\sin\theta'|+l\rangle \\ \cos\theta'\sin\theta'|-l\rangle + \sin^2\theta'|+l\rangle \end{pmatrix}$, and the transmission field from the $BS_2$ is expressed as: $\frac{1}{2\sqrt{2}} \begin{pmatrix} \cos(\theta'-\phi')\cos\theta'[\cos\phi'|+l\rangle + \sin\phi'|-l\rangle] \\ \cos(\theta'-\phi')\sin\theta'[\cos\phi'|+l\rangle + \sin\phi'|-l\rangle] \end{pmatrix}$. We consider the Jones matrix $\frac{1}{\sqrt{2}}\begin{pmatrix} 1 & 1 \\ 1 & -1 \end{pmatrix}$ and the influence of reflection on the polarization and OAM, the output field from the $BS_3$ can be expressed as:

$$E_T = \frac{1}{4} \begin{pmatrix} [-\cos^2\theta' + \cos(\theta'-\phi')\cos\theta'\cos\phi']|-l\rangle + [-\cos\theta'\sin\theta' + \cos(\theta'-\phi')\cos\theta'\sin\phi']|+l\rangle \\ [\cos\theta'\sin\theta' - \cos(\theta'-\phi')\sin\theta'\cos\phi']|-l\rangle + [\sin^2\theta' - \cos(\theta'-\phi')\sin\theta'\sin\phi']|+l\rangle \end{pmatrix}$$

(A5)

Then, $I_T = |E_T|^2 = \frac{1}{16}[1 - \cos^2(\theta'-\phi')] = \frac{1}{16}\sin^2(\theta'-\phi')$.

After $I_1'$, $I_2'$ and $I_T$ have been obtained, the normalized probabilities for the polarization and the OAM states $P_{P,O}(\theta',\phi')$ can be expressed as $P_{P,O}(\theta',\phi') = \frac{(2I_T - I_2' - I_1')^2}{I * I_2'}$ [33], here $I$ represents the input total intensity, $\theta'$ and $\phi'$ still represent polarization and OAM rotated angles. Compared with the expression in the Sec.II, we use $\theta'$ and $\phi'$ instead of $\theta$ and $\phi$ because they come from different elements in the experiments. Then, $P_{H,+l} = P_{P,O}(\theta',\phi')$, $P_{H,-l} = P_{P,O}(\theta',\phi'+\pi/2)$, $P_{V,+l} = P_{P,O}(\theta'+\pi/2,\phi')$ and $P_{V,-l} = P_{P,O}(\theta'+\pi/2,\phi'+\pi/2)$. The correlation function $C(\theta',\phi') = P_{H,+l} - P_{H,-l} - P_{V,+l} + P_{V,-l} = \cos[2(\theta'-\phi')]$ and the CHSH measurement B can also be obtained. The calculated results are shown in Fig.3.

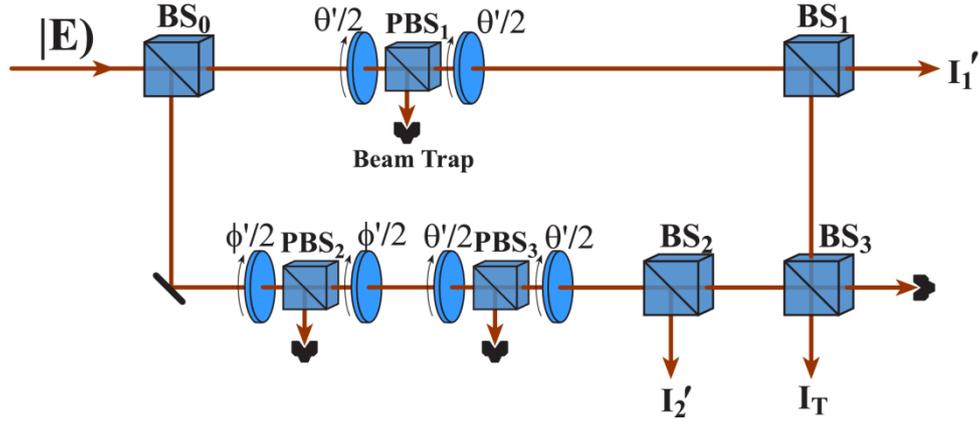

**Fig.A1** Experimental setup for the CHSH-type Bell's measurement of the correlation between polarization and OAM with any mode number.

# References

[1] J. S. Bell, *On the Einstein-Podolsky-Rosen paradox*, Physics **1**, 195 (1964).

[2] S. J. Freedman and J. F. Clauser, *Experimental Test of Local Hidden-Variable Theories*, Phys. Rev. Lett. **28**, 938 (1972).

[3] A. Aspect, J. Dalibard, and G. Roger, *Experimental Test of Bell's Inequalities Using Time-Varying Analyzers,* Phys. Rev. Lett. **49**, 1804 (1982).

[4] G. Weihs, T. Jennewein, C. Simon, H. Weinfurter, and A. Zeilinger, *Violation of Bell's Inequality under Strict Einstein Locality Conditions*, Phys. Rev. Lett. **81**, 5039 (1998).

[5] M. A. Rowe, D. Kielpinski, V. Meyer, C.A. Sackett, W. M. Itano, C. Monroe, and D. J. Wineland, *Experimental Violation of a Bell's Inequality with Efficient Detection*, Nature (London) **409**, 791 (2001).

[6] K. H. Kagalwala, G. Di Giuseppe, A. F. Abouraddy, and B. E. A. Saleh, *Bell's measure in classical optical coherence*, Nat. Photonics. **7**, 72 (2013).

[7] C. V. S. Borges, M. Hor-Meyll, J. A. O. Huguenin, and A. Z. Khoury, *Bell-like inequality for the spin-orbit separability of a laser beam*, Phys. Rev. A **82**, 033833 (2010).

[8] K. F. Lee and J. E. Thomas, *Experimental Simulation of Two-Particle Quantum Entanglement using Classical Fields*, Phys. Rev. Lett. **88**, 097902 (2002).

[9] M. A. Goldin, D. Francisco, and S. Ledesma, *Simulating Bell Inequality Violations with Classical Optics Encoded Qubits*, J. Opt. Soc. Am. B **27**, 779 (2010).

[10] B. N. Simon, S. Simon, F. Gori, M. Santarsiero, R. Borghi, N. Mukunda, and R. Simon, *Nonquantum Entanglement Resolves a Basic Issue in Polarization Optics*, Phys. Rev. Lett. **104**, 023901 (2010).

[11] X. F. Qian and J. H. Eberly, *Entanglement and Classical Polarization States*, Opt. Lett. **36**, 4110 (2011).

[12] S. K. Goyal, F. S. Roux, A. Forbes, and T. Konrad, *Implementing Quantum Walks Using Orbital Angular Momentum of Classical Light*, Phys. Rev. Lett. **110**, 263602 (2013).

[13] P. Chowdhury, A. S. Majumdar, and G. S. Agarwal, *Nonlocal Continuous-Variable Correlations and Violation of Bell's Inequality for Light Beams with Topological Singularities*, Phys. Rev. A **88**, 013830 (2013).

[14] R. J. C. Spreeuw, *A Classical Analogy of Entanglement*, Found. Phys. **28**, 361 (1998).

[15] P. Ghose and A. Mukhrjee, *Entanglement in Classical Optics*, Reviews in Theoretical Science **2**, 274 (2014).

[16] R. J. C. Spreeuw, *Classical Wave-Optics Analogy of Quantum-Information Processing*, Phys. Rev. A **63**, 062302 (2001).

[17] D. Francisco and S. Ledesma, *Classical Optics Analogy of Quantum Teleportation*, J. Opt. Soc. Am. B **25**, 383 (2008).

[18] A. Luis, *Coherence, Polarization, and Entanglement for Classical Light Fields*, Opt. Commun. **282**, 3665 (2009).

[19] F. Töppel, A. Aiellol, C. Marquardt, E. Giacobino, and G. Leuchs, *Classical Entanglement in Polarization Metrology*, New J. Phys. **16**, 073019 (2014).

[20] L. Allen, M. W. Beijersbergen, R. J. C. Spreeuw, and J. P. Woerdman, *Orbital Angular Momentum of Light and the Transformation of Laguerre-Gaussian Laser Modes*, Phys. Rev. A **45**, 8185 (1992).

[21] A. Mair, A. Vaziri, G. Weihs, and A. Zeilinger, *Entanglement of the Orbital Momentum States of Photons*, Nature, **412**, 313 (2001).

[22] J. Wang, J.-Y. Yang, I. M. Fazal, N. Ahmed, Y. Yan, H. Huang, Y.-X. Ren, Y. Yue, S. Dolinar, M. Tur, and A. E. Willner, *Terabit Free-Space Data Transmission Employing Orbital*

*Angular Momentum Multiplexing*, Nat. Photonics. **6**, 488 (2012).

[23] A. M. Yao, M. J. Padgett, *Orbital Angular Momentum: Origins, Behavior and Applications*, Adv. Opt. Photon. **3**, 161 (2011).

[24] Y. S. Weinstein, M. A. Pravia, E. M. Fortunato, S. Lloyd, and D. G. Cory, *Implementation of the Quantum Fourier Transform*, Phys. Rev. Lett. **86**, 1889 (2001).

[25] M. O. Scully and M. S. Zubairy, *Cavity QED Implementation of the Discrete Quantum Fourier Transform*, Phys. Rev. A **65**, 052324 (2002).

[26] J. F. Zhang, G. L. Long, Z. W. Dheng, W. Z. Liu, and Z. H. Lu, *Nuclear Magnetic Resonance Implementation of a Quantum Clock Synchronization Algorithm*, Phys. Rev. A **70**, 062322 (2004).

[27] J. Chiaverini, J. Britton, D. Leibfried, E. Knill, M. D. Barrett, R. B. Blakestad, W. M. Itano, J. D. Jost, C. Langer, R. Ozeri, T. Schaetz, D. J. Wineland, *Implementation of the Semiclassical Quantum Fourier Transform in a Scalable System*, Science **308**, 997 (2005).

[28] R. Barak and Y. Ben-Aryeh, *Quantum Fast Fourier Transform and Quantum Computation by Linear Optics*, J. Opt. Soc. Am. B **24**, 231 (2007).

[29] H. F. Wang, X. Q. Shao, Y. F. Zhao, S. Zhang, and K. H. Yeon, *Protocol and Quantum Circuit for Implementing the N-bit Discrete Quantum Fourier Transform in Cavity QED*, J. Phys. B **43**, 065503 (2010).

[30] H. F. Wang, A. D. Zhu, S. Zhang, and K. H. Yeon, *Simple Implementation of Discrete Quantum Fourier Transform Via Cavity Quantum Electrodynamics*, New J. Phys. **13**, 013021 (2011).

[31] M.W. Beijersbergen, L. Allen, H. E. L. O. van der Veen, and J. P. Woerdman, *Astigmatic laser mode converters and transfer of orbital angular momentum,* Opt. Commun. **96**, 123 (1993).

[32] M. A. Nielsen and I. L. Chuang, *Quantum computation and quantum information*, Cambridge University Press (2000).

[33] X-F Qian, Bethany Little, John C. Howell, and J. H. Eberly, *Violation of Bell's Inequalities with Classical Shimony-Wolf States: Theory and Experiment*, arXiv preprint arXiv:1406.3338 (2014).